\documentclass[aps,prb,a4paper,twocolumn,showpacs]{revtex4}

\usepackage{graphicx}
\usepackage{epstopdf}
\usepackage{amsmath}
\usepackage{amsfonts}
\usepackage{amssymb}
\usepackage{color}

\makeatother
\begin{document}
\title[]{The influence of antichiral edge states on Andreev reflection in graphene-superconductor junction}

\author{Chao Wang,$^{1,2}$ Lin Zhang,$^{1}$ Peipei Zhang,$^{1}$ Juntao Song,$^{1*}$ Yu-Xian Li$^{1*}$}

\affiliation{$^{1}$College of Physics and Hebei Advanced Thin Film Laboratory,
Hebei Normal University, Shijiazhuang 050024, People's Republic of China\\
$^{2}$College of Physics, Shijiazhuang University, Shijiazhuang 050035, People's Republic of China}

\vspace{10pt}

\begin{abstract}
Using the tight-binding model and the non-equilibrium Green's function method, we study Andreev reflection in a  graphene-superconductor junction, where graphene has two nonequal Dirac cones split in energy and therefore time-reversal symmetry is broken. Due to the antichiral edge states of the current graphene model, an incident electron traveling along the edges makes a distinct contribution to Andreev reflections. In a two-terminal device, because  Andreev retroreflection is not allowed for just the antichiral edges, in this case the mutual scattering between edge and bulk states is necessary, which leads to the coefficient of Andreev retroreflection always being symmetrical about the incident energy. In a four-terminal junction, however, the edges are parallel to the interface of superconductor and graphene, so at the interface an incident electron traveling along the edges can be retroreflected as a hole into bulk modes, or specularly reflected as a hole into antichiral edge states again. It is noted that the coefficient of specular Andreev reflection keeps symmetric as to the incident energy of electrons, which is consistent with the reported results before; however, the coefficient of Andreev retroreflection shows an unexpected asymmetrical behavior due to the presence of antichiral edge states. Our results present some ideas to study the antichiral edge modes and Andreev reflection for a graphene model with the broken time-reversal symmetry.
\end{abstract}

\date{\today}

\maketitle

\narrowtext
\section{Introduction}

Andreev reflection ~\cite{Andreev,deJong} is related to a particle-tunneling process on the interface of conductor and superconductor, where an incident electron of the conductor is reflected as a hole into the conductor and simultaneously a Cooper pair is formed in the superconductor. When the bias between conductor and superconductor is small enough, the transport property of the conductor-superconductor junction is mainly determined by Andreev reflection. Therefore, Andreev reflection is always a widely studied issue in the field of condensed-matter physics. In the conventional Andreev reflection, the incident electron and the reflected hole come from the same band (conduction band or valence band). This is why the reflected hole holds the opposite transport direction of the incident electron. In general, the conventional Andreev reflection is called Andreev retroreflection or intraband reflection.

\begin{figure}
\centering
  \includegraphics[width=1.0\columnwidth]{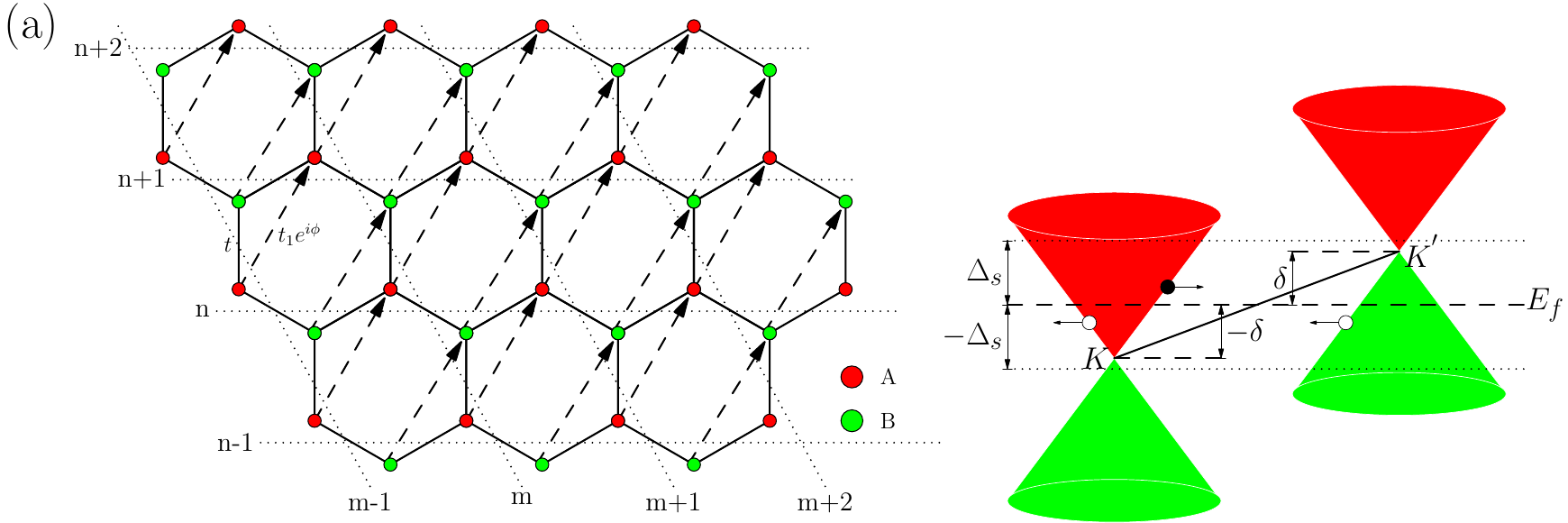}\\[10pt]
  \hspace{0.0\linewidth}
  {\includegraphics[width=0.45\columnwidth]{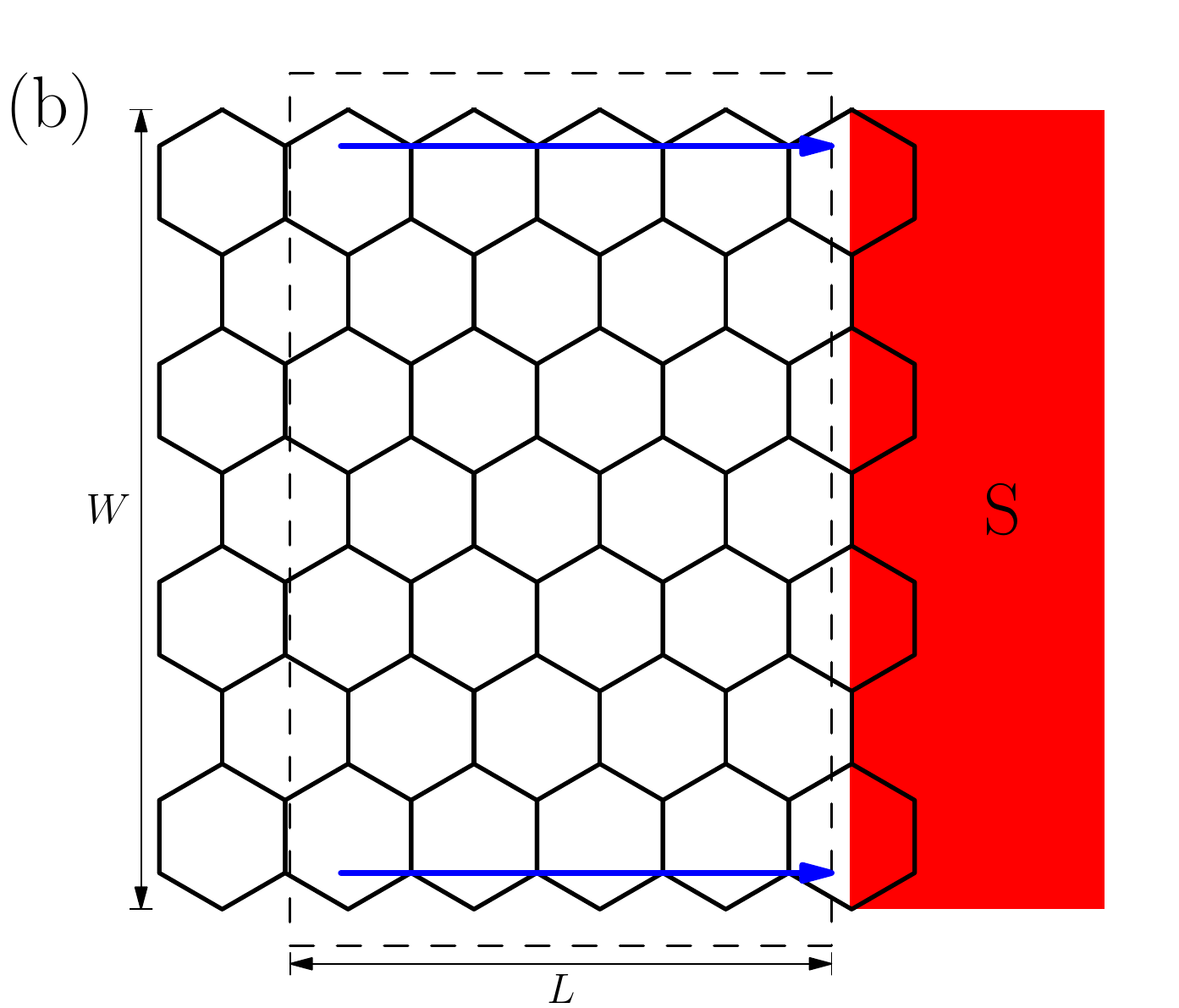}}
   \hspace{0.0\linewidth}
  {\includegraphics[width=0.5\columnwidth]{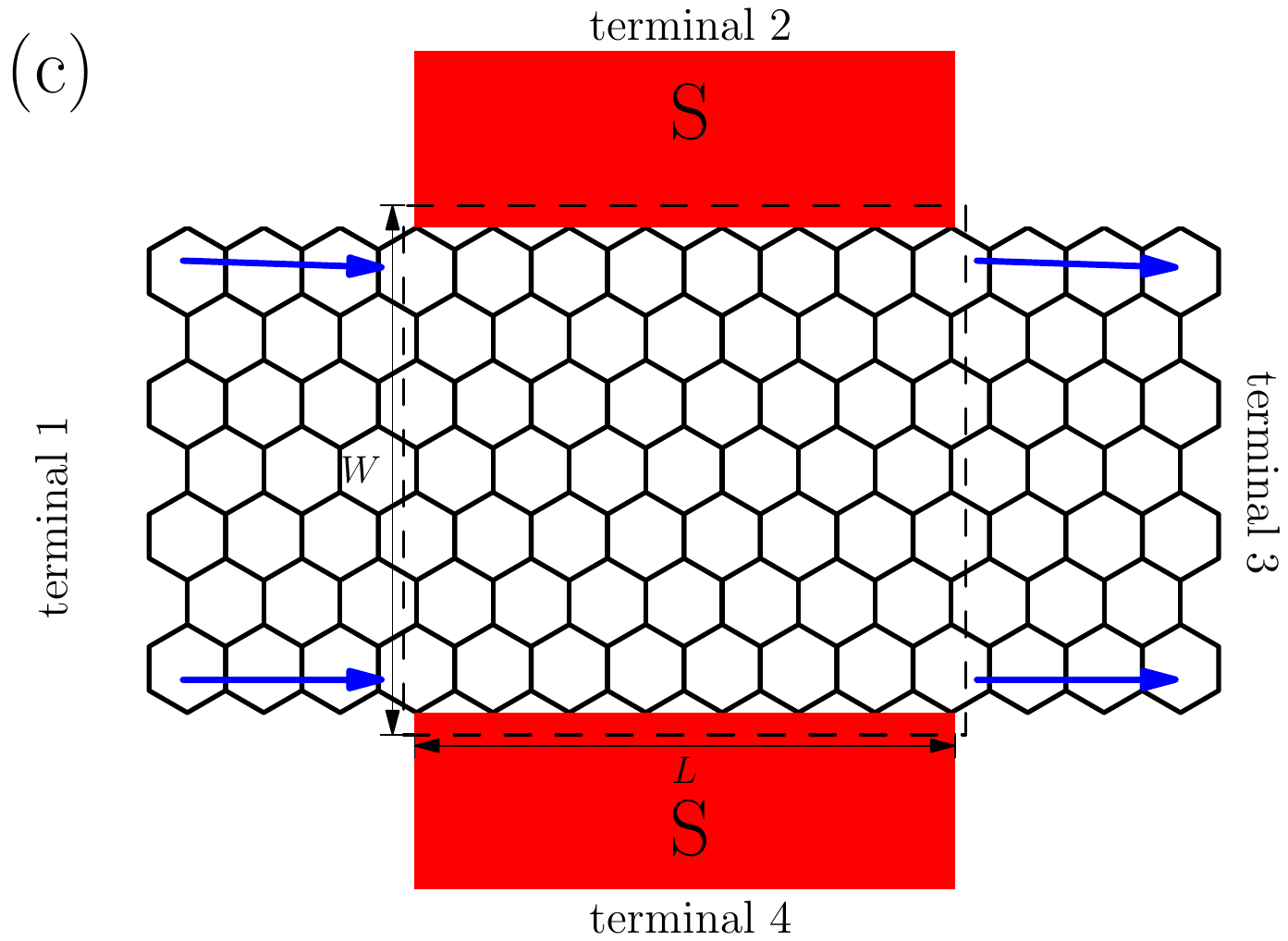}}
\caption{(Color online) (a) The zigzag graphene ribbon with the split hopping term (left) and the schematic diagram of the energy band (right). In the right diagram, $E_{f}$ is Fermi energy and $\Delta_{s}$ represents the superconductor gap. Dirac cones are split in energy by $\delta=\sqrt{3}t_{1}$ when $\phi=\pi/2$. The black solid dot stands for the incident electron and the hollow dots stand for the reflected holes in $K$ and $K'$ valleys, respectively. The black solid line connecting two Dirac points is the edge modes. (b) and (c) are the schematic diagrams for the two-terminal and four-terminal graphene-superconductor junction, respectively. The blue lines represent the antichiral edge states with the same dispersion. $L$($W$) is the length (width) of the center region of the graphene-superconductor junction. $L=4$($L=8$) in the two-terminal (the four-terminal) device and $W=4$ in (b) and (c). }\label{fig1}
\end{figure}

After successfully producing the graphene in experiment \cite{Novoselov}, the physical characteristics of graphene ~\cite{Neto,Novoselov2005,Geim,Sarma,Haldane,Cheianov,Young,Nakada} became a research focus in condensed-matter physics. In the graphene-superconductor junction, besides Andreev retroreflection, there exists another kind of Andreev reflection, which is called specular Andreev reflection, due to the gapless energy band of graphene. Unlike Andreev retroreflection, the reflected hole travels along the specular path of the incident electron in specular Andreev reflection. Especially, the specular-reflected hole and the incident electron belong to the different bands, which makes the specular Andreev reflection to also be named interband Andreev reflection. For example, if an incident electron comes from the conduction band, the reflected hole must enter into the valence band in the specular Andreev reflection. Ever since, many research groups have studied the electron transport properties in graphene-superconductor hybrid systems ~\cite{Beenakker,Amet2016,Rickhaus,Titov,Schelter,Xie2009,Efetov,Linder,Cheng,Yanxia, Xing,Rainis,Sun2017}.

For the band structure of a graphene nanoribbon with zigzag edges, valence and conduction bands touch each other at two points ($K$ and $K'$), which are called Dirac points. Due to the time-reversal symmetry, the incident electron and the reflected hole come from the different valleys in Andreev reflection. But in some conditions, the hole can be reflected into the same valley as the incident electron, so there are many works ~\cite{Zhou,Niu2007,Beenakker2007,WangYao2012,Geim2014,Niu2008,SongLD2012,SongVP2013} to study manipulating the valley index for designing the electronic device. As we know, for the pristine graphene ribbon, there are zero-energy flat bands connecting the two Dirac points, which are dispersionless. In previous works ~\cite{Haldane,mele2005,kane2005,kane2010,Song2016}, the edge modes behave as not only  chiral states with the quantum Hall effect, but also helical states with the quantum spin Hall effect. When a second-neighbor hopping term is included along a certain direction in the  Hamiltonian ~\cite{Aoki2010,Franz2018}, which is called the split hopping term in this paper, the two Dirac points are split in energy along opposite directions [see Fig. \ref{fig1}(a)]. Because of this, both of the edge modes acquire the same dispersion, namely the same movement velocity, which therefore means the breakdown of the time-reversal symmetry.

\begin{figure}[tbp]
\centering
\includegraphics[width=1.0\columnwidth,angle=0]{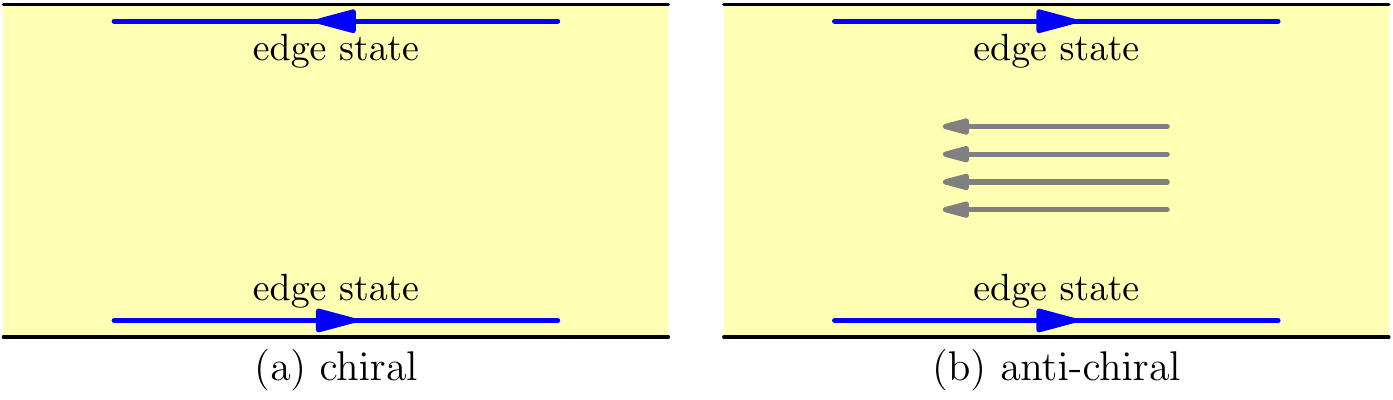}
\caption{(Color online) The schematic diagram for (a) chiral edge states and (b) antichiral edge states in the ribbon model.  } \label{fig2}
\end{figure}

As is known, a chiral edge state moves forward by encircling the sample boundary, so in a ribbon sample the movement directions of a chiral edge state on two edges  are always opposite to each other [see Fig. \ref{fig2}(a)]. Given the movement directions of edge states are the same on two edges of a ribbon sample [see Fig. \ref{fig2}(b)], it is generally called an antichiral edge state, proposed in the modified Haldane model recently \cite{Franz2018}. The antichiral edge modes are topologically protected by the chiral symmetry. Besides the same dispersion of the edge states, the bulk modes have opposite moving directions to edge states in the model. Due to intriguing characteristics of the antichiral edge states, it arouses the concern of researchers.

In this paper, we study  Andreev reflection in the graphene-superconductor junction, where the graphene nanoribbon has antichiral edge states. Different from the cases reported in previous works ~\cite{Xie2009,Cheng,zhang2015,Cheng2011,wang2018,Alidoust2017,Zareyan2012}, in our model both edge states of the ribbon acquire the same velocity and dispersion. Noted that, due to the broken time-reversal symmetry of antichiral edge states \cite{vila2019}, the reflected hole can get into both Dirac valleys, in theory. In a two-terminal junction, when the energy of the Dirac points $\varepsilon_{0}$ is set to be zero, the Andreev reflection coefficient is nonzero only for the incident energy $|E|<\delta$, where $\delta$ is the shifted energy of the Dirac point. When $\varepsilon_{0}$ is nonzero, the Andreev reflection coefficient is nonzero only for $|E|<|\delta-\varepsilon_{0}|$. In a four-terminal junction, the influence of the antichiral edge states on two kinds of Andreev reflection is also studied. It is found that incident electrons which travel along the edges of the ribbon are mainly specularly reflected as holes at the interface between graphene and superconductor. This conclusion is gradually destroyed with increasing the on-site energy $\varepsilon_0$. Especially, asymmetrical Andreev retroreflection coefficients about the incident energy are obtained.

The rest of this paper is arranged as follows. In Sec. \uppercase\expandafter {\romannumeral 2}, the model Hamiltonian for the system is presented and the formulas for calculating  the Andreev reflection coefficients are derived. Our main results are shown and discussed in Sec. \uppercase \expandafter {\romannumeral 3}. Finally, a brief conclusion is presented in Sec. \uppercase\expandafter {\romannumeral 4}.\\

\section{Model and Formula}

The graphene-superconductor junction investigated here is related to a graphene ribbon with split Dirac cones [see Fig. \ref{fig1}(a)] and superconductor terminals connected to the graphene ribbon [see Figs. \ref{fig1}(b) and \ref{fig1}(c)]. The total Hamiltonian of this junction can be represented as
\begin{equation}\
 H=H_{G}+H_{S}+H_{T},
\end{equation}
where $H_{G}$, $H_{S}$, and $H_{T}$ are the Hamiltonians of the graphene ribbon, superconductor terminals, and the coupling between the
graphene ribbon and superconductor terminals, respectively.

In the tight-binding representation, $H_{G}$ is given by \cite{Neto}
\begin{eqnarray}\label{HamiltonianTotal}
&H_{G}=\sum_{m,n}\varepsilon_{0}[a^{\dag}_{\emph{m,n}}a_{\emph{m,n}}+b^{\dag}_{\emph{m,n}}b_{\emph{m,n}}]\qquad \qquad    \nonumber\\
&-t[a_{\emph{m,n}}^{\dagger}b_{\emph{m,n}}+a_{\emph{m,n}}^{\dagger}b_{\emph{m-1,n}}+a_{\emph{m,n}}^{\dagger}b_{\emph{m,n-1}}+\mathrm{H.c.}]\nonumber\\
&-t_{1}[e^{\mathrm{i}\phi}(a_{\emph{m,n}}^{\dagger}a_{\emph{m+1,n+1}}+b_{\emph{m,n}}^{\dagger}b_{\emph{m+1,n+1}})+\mathrm{H.c.}],
\end{eqnarray}
where $a_{\emph{m,n}}^{\dagger}$ ($a_{\emph{m,n}}$) and $b_{\emph{m,n}}^{\dagger}$ ($b_{\emph{m,n}}$) are the creation (annihilation) operators  of the sublattices A and B at site $(\emph{m,n})$. $\varepsilon_{0}$ is the onsite energy, which corresponds to the position of Dirac points for pristine graphene. The second term in Eq. (\ref{HamiltonianTotal}) stands for the nearest-neighbor-hopping Hamiltonian. The third term is the split hopping Hamiltonian in the graphene ribbon [see Fig. \ref{fig1}(a)]. We consider that the graphene region is directly coupled to the superconductor terminal. Described by a continuum model, the superconductor terminal is represented by a BCS Hamiltonian,
\begin{eqnarray}
H_{S}=\sum_{\textbf{\emph{k}},\sigma}\varepsilon_{\textbf{\emph{k}}}C^{\dagger}_{\textbf{\emph{k}}\sigma}C_{\textbf{\emph{k}}\sigma}
+\sum_{\textbf{\emph{k}}}(\Delta C_{\textbf{\emph{k}}\downarrow}C_{-\textbf{\emph{k}}\uparrow}
+\Delta^{*}C^{\dagger}_{-\textbf{\emph{k}}\uparrow}C^{\dagger}_{\textbf{\emph{k}}\downarrow}),
\end{eqnarray}
where $\Delta=\Delta_{s}e^{\mathrm{i}\theta}$. Here $\Delta_{s}$ is the superconductor gap and $\theta$ is the superconductor phase. The coupling between superconductor terminal and graphene is described by
\begin{eqnarray}
H_{T}=-\sum_{\emph{m,n},\sigma}\emph{t}_c(a^{\dagger}_{\emph{m,n},\sigma}+b^{\dagger}_{\emph{m,n},\sigma})C_{\sigma}(x_{\emph{i}})+\mathrm{H.c.}
\end{eqnarray}
Here $x_{\emph{i}}$ and $(m,n)$ represent the positions of the coupling atoms on the interface of superconductor and graphene, and $C_{\sigma}(x)=\sum_{\emph{k}_{x},\emph{k}_{y}}e^{\mathrm{i}k_{x}x}C_{\textbf{\emph{k}},\sigma}$ is the annihilation operator in real space. Note that $\sigma$ represents the spin index and $\emph{t}_c$ is the coupling strength between graphene and superconductor terminals.

We now turn to analyze the process that an incident electron from the graphene terminal is reflected into a hole with a Cooper pair emerging in the superconductor terminal. Using nonequilibrium Green's function method, we can calculate the retarded and advanced Green's function $G^{r}(E)=[G^{a}]^{\dagger}=1/(EI-H_{C}-\sum_{\alpha}\mathbf{\Sigma}^{r}_{\alpha})$, where $H_{C}$ is the Hamiltonian of the center region in the Nambu representation and $I$ is the unit matrix with the same dimension as $H_{C}$. The center region is the rectangular region surrounded by the dashed line in Figs. \ref{fig1}(b) and 1(c). $\mathbf{\Sigma}_{\alpha}^{r}=t_c g_{\alpha}^{r}(E)t_c$ is the retarded self-energy due to the coupling to the terminal $\alpha$,  where $g_{\alpha}^{r}(E)$  is the surface Green's function of the terminal $\alpha$. We can numerically calculate the surface Green's function of the graphene terminals. For superconductor terminals, the surface Green's function ~\cite{Xie2009,wang2018,Sun2011,Sun2016,Song2017} in real space is
\begin{eqnarray}\label{1}
 g_{\alpha,ij}^{r}(E)=i\pi\rho\beta(E)J_{0}[k_{f}(x_{i}-x_{j})]\left(
\begin{array}{cc}
  1 & \Delta/E \\
  \Delta^{*}/E & 1 \\
\end{array}
\right),
\end{eqnarray}
where $\alpha=2,4,$ and $\rho$ is the density of normal electron states. $J_{0}[k_{f}(x_{i}-x_{j})]$ is the Bessel function of the first kind with the Fermi wave vector $k_{f}$. $\beta(E)=-iE/\sqrt{\Delta_{s}^{2}-E^2}$ for $|E|<\Delta_{s}$ and $\beta(E)=|E|/\sqrt{E^2-\Delta_{s}^{2}}$ for $|E|>\Delta_{s}$.

The Andreev reflection  coefficients for the incident electron coming from the graphene terminal 1 can  be obtained ~\cite{Xie2009,zhang2015},
\begin{eqnarray}\label{AndRfC}
&T_{A,11}(E)=\mathrm{Tr}\{\Gamma_{1,\uparrow\uparrow}G^{r}_{\uparrow\downarrow}\Gamma_{1,\downarrow\downarrow}G^{a}_{\downarrow\uparrow}\},\nonumber\\
&T_{A,13}(E)=\mathrm{Tr}\{\Gamma_{1,\uparrow\uparrow}G^{r}_{\uparrow\downarrow}\Gamma_{3,\downarrow\downarrow}G^{a}_{\downarrow\uparrow}\},
\end{eqnarray}
where the subscripts $\uparrow\uparrow$, $\downarrow\downarrow$, $\uparrow\downarrow$ and $\downarrow\uparrow$ represent the 11, 22, 12, and 21 matrix elements, respectively, in the Nambu representation. The linewidth function $\Gamma_{\alpha}$ is defined with the aid of self-energy as $\Gamma_{\alpha}=i[\mathbf{\Sigma}_{\alpha}^{r}-(\mathbf{\Sigma}_{\alpha}^{r})^{\dag}]$. $T_{A,11}$ and $T_{A,13}$ represent the coefficients of  Andreev retroreflection and specular Andreev reflection, respectively. Because there is only one graphene terminal for the two-terminal junction in this paper, the coefficient of Andreev reflection $T_{A,11}$ is written as $T_A$ for simplicity. Because the Andreev reflection from an electron to a hole is equivalent to that from a hole to an electron under particle-hole symmetry, in this paper we only consider the Andreev reflections, where an incident electron is reflected as a hole. \\

\section{Results and discussion}
In numerical calculations, we set the nearest-neighbor-hopping energy $t=2.75 eV$.  The length of the nearest-neighbor C-C bond is set to be $a_{0}=0.142$ $\mathrm{nm}$ as in a real graphene sample. The superconductor gap is set to be $\Delta_{s}=0.02t$, the superconductor phase $\theta=0$ and the Fermi wave vector $k_{f}=10$ $\mathrm{nm^{-1}}$. For the convenience of discussing the influence of the split hopping term, the Fermi energy $E_{f}$ is always set to be zero and the hopping phase $\phi$ in the split hopping term takes the value $\pi/2$.

\begin{figure}[tbp]
\centering
\includegraphics[width=1.0\columnwidth,angle=0]{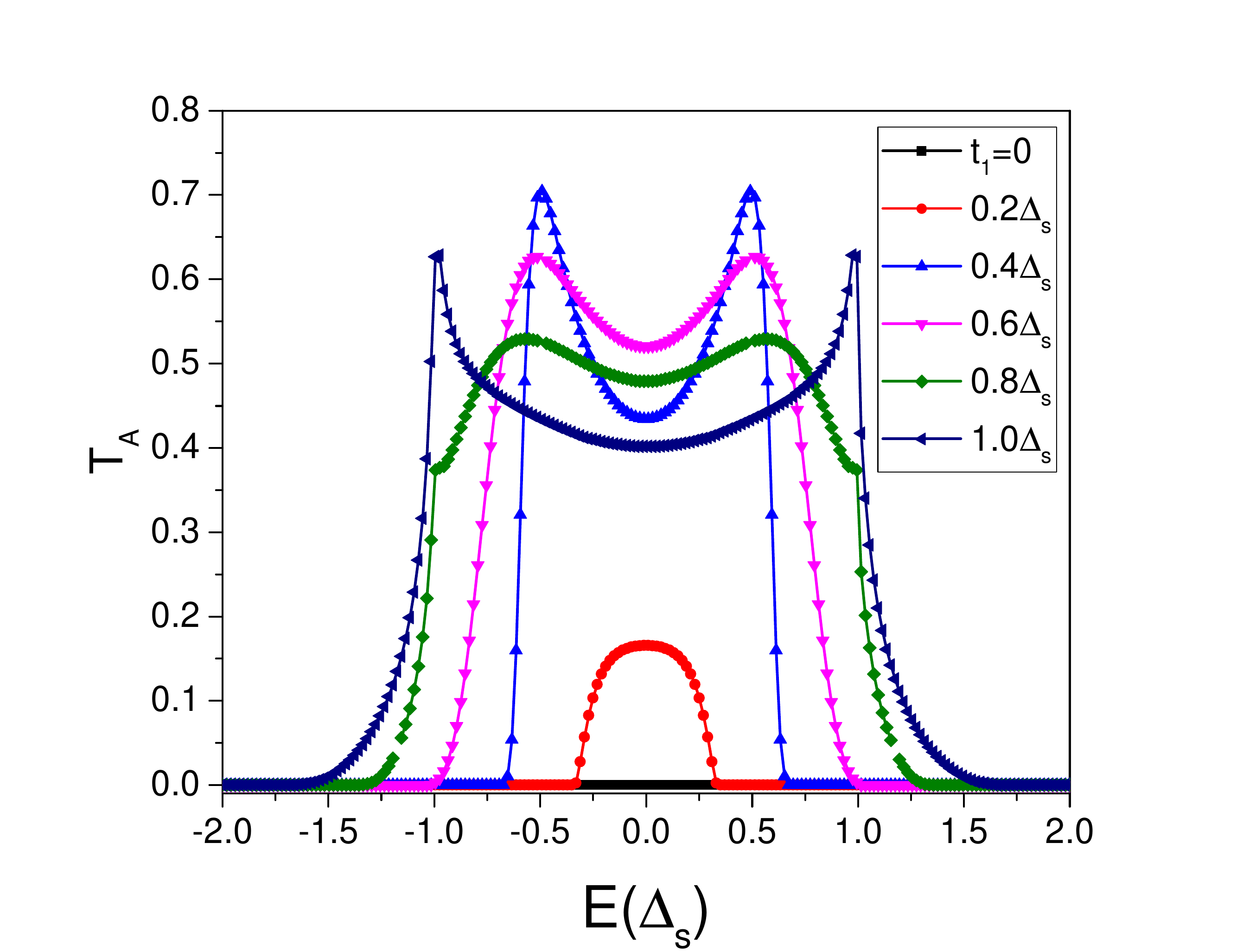}
\caption{(Color online) In the two-terminal graphene-superconductor junction $T_A$ vs the incident energy $E$ for different split hopping strength $t_{1}$. Here, $\varepsilon_0=0$, the width of the central region $W=80$, and the length of the central region $L=60$. } \label{fig3}
\end{figure}

Figure \ref{fig3} shows the Andreev reflection coefficient $T_A$ as a function of the incident energy $E$ for the different split hopping strength $t_1$ in the two-terminal graphene-superconductor junction [see Fig. \ref{fig1}(b)], where $\varepsilon_{0}=0$.  In the two-terminal junction, the reflected hole can only travel back into the left graphene terminal, no matter which kind of Andreev reflection occurs. Due to the normal incidence of electron to the interface, the momentum component paralleling to the interface could be supposed to be zero. This supposition always keeps true, no matter the incident electron travels along the edges or within the bulk. So when Andreev reflections take place in the two-terminal junction, the momentum of the incident electron can only be changed in the direction perpendicular to the interface, which exactly corresponds to the case of  Andreev retroreflection. Namely, specular Andreev reflection is largely restrained in the two-terminal junction.

When $t_1=0$, the sample is exactly the pristine graphene ribbon with zigzag edges. Undoubtedly, from the perspective of band structure, only the interband reflection is permitted at $\varepsilon_{0}=0$, which corresponds to the case of the specular Andreev reflection. In  Fig. \ref{fig3}, it can be therefore obtained that $T_A$ is zero for $t_1=0$. In a word, both  Andreev retroreflection and  specular Andreev reflection should be zero in the two-terminal pristine graphene-superconductor junction \cite{Xie2009}, due to the combined restrictions of the two-terminal junction and band structure.

The above situation changes immediately for a nonzero split hopping term $t_1$, where two Dirac cones are split in energy and thus time-reversal symmetry is broken now. As shown in Fig. \ref{fig3}, $T_A$ shows nonzero values for nonzero $t_1$. When $t_1$ takes a small value $0.2\Delta_s$, the coefficient of $T_{A}$ still keeps zero for $|E|>\delta$, where $\delta=\sqrt{3}t_1$ represents the energy difference between the present Dirac cones and the pristine ones. It is easy to understand from the fact that specular Andreev reflection is forbidden in the two-terminal junction and thus the interband reflection for $|E|>\delta$ should be zero. Consequently, it can be concluded that the nonzero coefficient of $T_{A}$ for $|E|<\delta$ should be ascribed to the intraband reflection, namely Andreev retroflection.

When $t_1$ increases to $0.4\Delta_s$, the nonzero part of $T_{A}$ is mainly confined to the range of $|E|<\delta$, as discussed above. In addition, two symmetrical peaks appear at the boundaries of $T_A$. It is easy to verify that bulk states account for these peaks, which will be discussed in detail below. As $t_1$ rises further, the velocity of the two antichiral edge states becomes larger and no more bulk states take part in the Andreev reflection, which induces the reduction of its contribution to the density of states at a fix energy. Thus, the magnitude of $T_A$ becomes smaller as $t_1$ continues changing from $0.6\Delta_s$ to $1.0\Delta_s$. When the split energy $\delta$ of Dirac cones is larger than the superconductor gap $\Delta_s$,  the nonzero range of $T_A$ is mainly regulated by the superconductor gap. It is because there is no Andreev reflection for $|\Delta_{s}|<|E|<|\delta|$ according to the definition of Andreev reflection, which is a process of electron-hole conversion inside the superconductor gap.  To sum up, for the two-terminal hybrid junction above, there is mainly the intraband Andreev reflection, namely Andreev retroreflection, and the antichiral edge states play a key role on Andreev reflection.

\begin{figure}[tbp]
\centering
\includegraphics[width=1.0\columnwidth,angle=0]{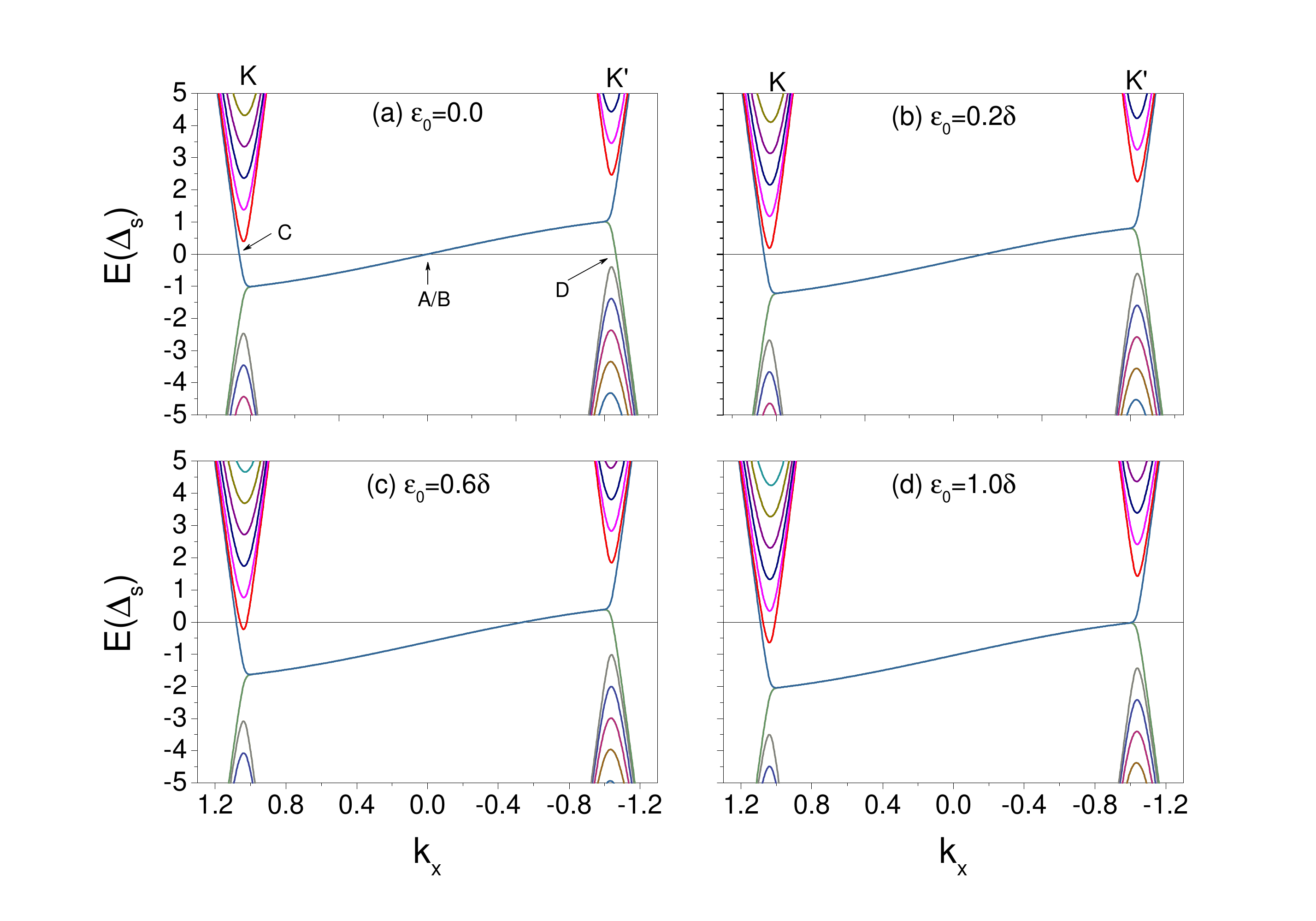}
\caption{(Color online) The band structures of the zigzag ribbon with the split Dirac cones. At $t_{1}=0.6\Delta_{s}$, the split energy of the Dirac point is $\delta=\sqrt{3}t_{1}=0.6\sqrt{3}\Delta_{s}\simeq1.0\Delta_{s}$. In (a)-(d), $\varepsilon_{0}=0$, $0.2\delta$, $0.6\delta$ and $1.0\delta$, respectively. The black solid line represents the Fermi energy. The width of the graphene ribbon is $W=80$.} \label{fig4}
\end{figure}

As is known, there are two kinds of Andreev reflection at the interface between pristine graphene and superconductor, Andreev retroreflection and specular Andreev reflection. Due to time-reversal symmetry, the incident electron and the reflected hole must come from different valleys in general. When the split hopping term is considered in our model, the low-energy effective Hamiltonian can be written as $H(k)=\delta\tau_z\otimes\sigma_0+k_x\tau_z\otimes\sigma_x+k_y\tau_0\otimes\sigma_y$ ~\cite{Beenakker,Aoki2010}, where the Pauli matrices $\tau_k$ and $\sigma_k$ represent the valley and sublattice indices, respectively. It is easy to verify that due to the nonzero split hopping term the time-reversal symmetry is broken, $\mathcal{T} H(k)\mathcal{T}^{-1}\neq H(-k)$, where $\mathcal{T}=i\tau_y\otimes\sigma_0\mathcal{C}$ is the time-reversal operator \cite{kane2005} with the complex operator $\mathcal{C}$. Without the constraint of time-reversal symmetry, an incident electron coming from the valley $K$ could be reflected as a hole into not only the other valley $K'$, but also the same valley $K$ [see Fig. \ref{fig1}(a)].

Besides, since the Dirac points of the adopted Hamiltonian are split in energy along the opposite directions,  two edge states of zigzag ribbon acquire the same dispersion and become anti-chiral. An electron traveling along the edge can be also reflected as a hole at the interface between graphene and superconductor. From Fig. \ref{fig3}, we can see that when the split hopping term is nonzero, the Andreev coefficient is nonzero for $|E|<\delta$ in the two-terminal junction. A question naturally arises as to what role the antichiral edge states plays in Andreev reflection.

For the sake of simplification, in the rest of the paper the value of the split hopping strength is fixed to be $t_{1}=0.6\Delta_{s}$, which results in the split of two Dirac points, $\delta=\sqrt{3}t_{1}\simeq1.0\Delta_{s}$.  To get a rudimentary understanding on the system, we first plot the band structures for different values of $\varepsilon_{0}$ in Fig. \ref{fig4}. As shown in Fig. \ref{fig4}(a) with $\varepsilon_{0}=0$, the two Dirac points are split along the opposite direction, distributing symmetrically on the two sides of the Fermi energy $E_F=0$. As increasing $\varepsilon_{0}$ gradually, the band structure moves down in whole and the inversion symmetry about the Fermi energy is also ruined, as seen in Figs. \ref{fig4}(b) and 4(c). It can be seen that a new bulk state at the $K$ point intersects with the Fermi energy for $\varepsilon_{0}=0.6\delta$ in Fig. \ref{fig4}(c). When $\varepsilon_{0}$ continues, increasing up to $\varepsilon_{0}=\delta$, the $K'$ point here goes back again to the Fermi energy exactly. No new bulk state appears around the Fermi energy in this case. Obviously, the on-site energy $\varepsilon_0$ can be used to tune the density of states near the Fermi energy.

\begin{figure}[tbp]
\centering
\includegraphics[width=1.0\columnwidth,angle=0]{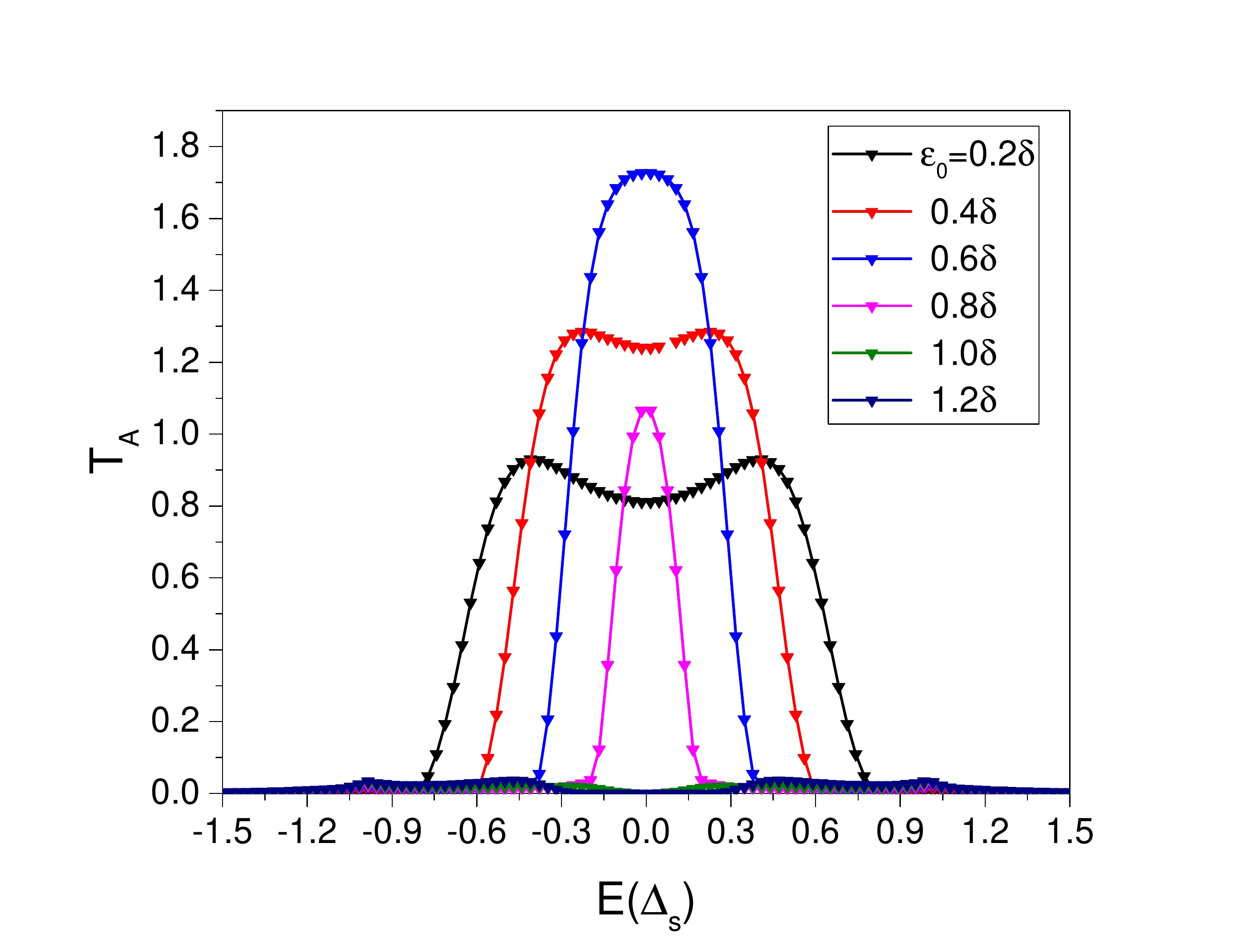}
\caption{(Color online) In the two-terminal graphene-superconductor junction, $T_A$ vs the incident energy $E$ for different $\varepsilon_{0}$ when $t_{1}=0.6\Delta_{s}$ and $\delta=\sqrt{3}t_{1}\simeq1.0\Delta_{s}$. The width and the length of the central region are $W=80$ and $L=60$, respectively.} \label{fig5}
\end{figure}

Next we calculate Andreev reflection coefficients with the incident energy $E$  for different values of $\varepsilon_{0}$ in the two-terminal junction. The curves are shown in Fig. \ref{fig5}. When $\varepsilon_{0}=0.2\delta$, two maximum values of $T_A$ up to $0.9$ can be observed within $|E|<0.8\Delta_{s}$. In this case, $T_A$ decays fast to zero for $|E|>0.8\Delta_{s}$. As increases to $0.4\delta$, the two symmetrical peaks of $T_A$ continue rising to $1.2$, and $T_A$ also reduces to almost zero for $|E|>0.6\Delta_{s}$. However, only one peak preserves at $\varepsilon_{0}=0.6\delta$. In addition, $T_A$ with nonzero value is confined to the region of $|E|<0.4\Delta_{s}$. As $\varepsilon_{0}$ takes the value $0.8\delta$, instead of increasing, $T_A$ presents an obvious drop in strength. Also, it becomes almost zero for  $|E|>0.2\Delta_{s}$. Unexpectedly, $T_A$ drops near to zero for $\varepsilon_{0}=1.0\delta$, even though the density of states is nonzero at Fermi energy. A similar situation can be observed at $\varepsilon_{0}=1.2\delta$.

Based on the detailed presentation and discussion above, here the main conclusions include four aspects: (1) In the case of $\varepsilon_0<\delta$, $T_A$ takes nonzero values within the region of $|E|<(\delta-\varepsilon_{0})$ and decays fast to zero for $|E|>(\delta-\varepsilon_{0})$.  The first reason is specular Andreev reflection is suppressed in the two-terminal junction. The second reason is the DOS of incident electron decreases for big $\varepsilon_0$. The third reason is the Andreev reflection coefficient is influenced by many factors, as the sample width and the  conservation principle of  momentum  and energy. (2) The two symmetrical peaks of $T_A$ for small $\varepsilon_{0}$ should be ascribed to the appearance of a few bulk states. (3) Due to the split hopping $t_1$, the time-reversal symmetry is broken and the edge states acquire the same dispersion in the graphene nanoribbon. The electrons of the antichiral edge states, for which the characteristics of $K$ and $K'$ Dirac points are mixed together, play an important role in Andreev reflection. The antichiral edge modes dominate the Andreev reflections near the energy of the Dirac points, which leads to nonzero values of $T_A$ for small $\varepsilon_{0}$. However, $T_A$ becomes almost zero when the antichiral edge modes are shifted away from the Fermi energy. These observations exactly illustrate that the broken time-reversal symmetry accounts for the nonzero coefficient of Andreev reflection $T_A$. (4) What should be particularly noted is that $T_A$ quickly becomes almost zero for $\varepsilon_{0}\geq\delta$, even there are a few edge and bulk states yet. In this sense, the role of the residual edge and bulk states in Andreev reflection is fundamentally different to that of the antichiral edge states due to the broken time-reversal symmetry.

\begin{figure}[tbp]
\includegraphics[width=1.0\columnwidth,angle=0]{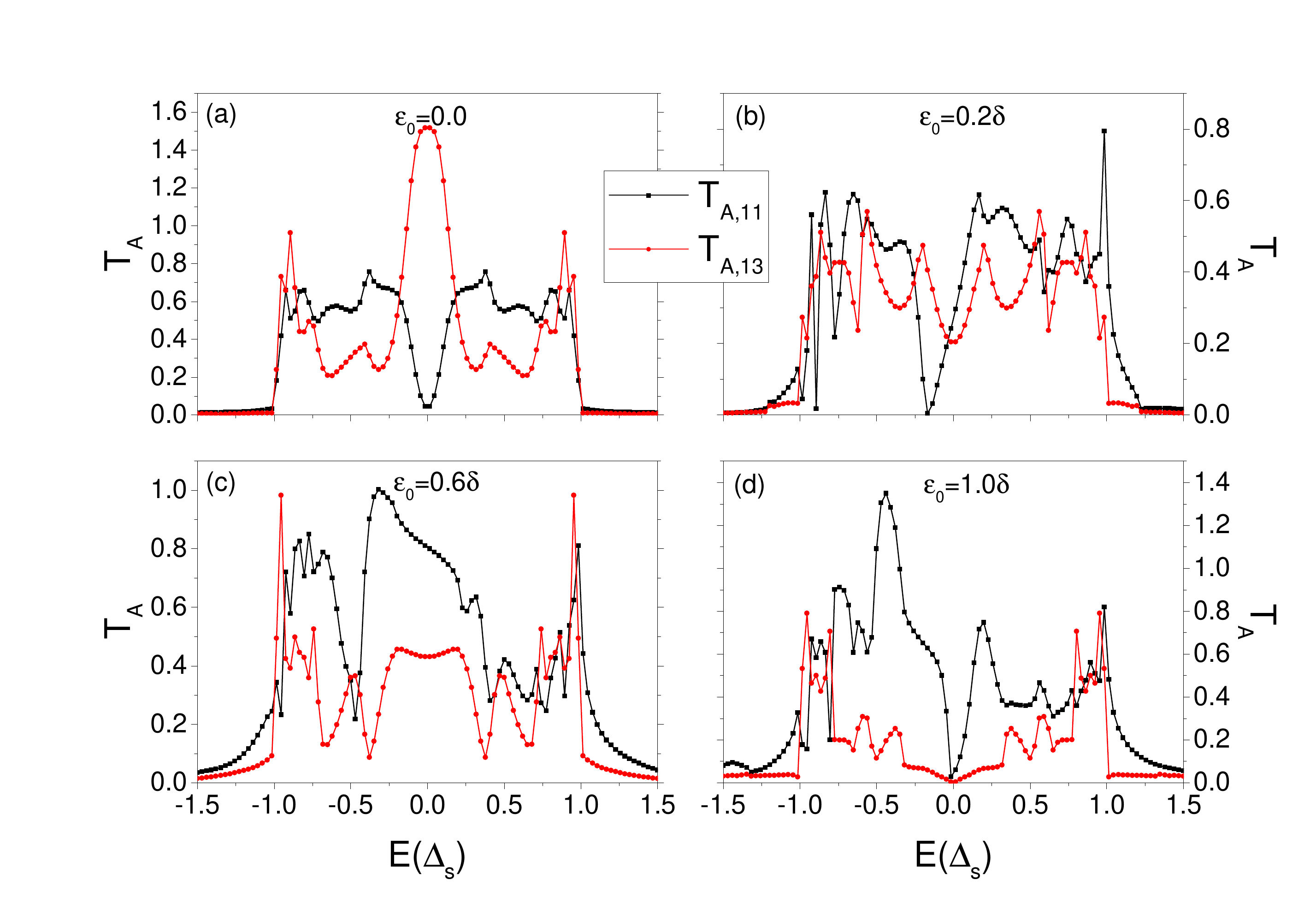}
\caption{(Color online) The coefficient of Andreev retro-reflection $T_{A,11}$ and the coefficient of specular Andreev reflection $T_{A,13}$ vs the incident energy $E$ in the four-terminal graphene-superconductor junction.  In (a)-(d), $\varepsilon_{0}$ takes $0.0\delta$, $0.2\delta$, $0.6\delta$, and $1.0\delta$, respectively. Here, the split strength $t_{1}=0.6\Delta_{s}$, the length of the center region $L=60$, and the width $W=80$.} \label{fig6}
\end{figure}

Andreev reflection in the two-terminal junction with the antichiral edge states is studied above. It is shown that the incident electron traveling along the edges can be retroreflected as a hole at the interface of the two-terminal junction. Due to only one graphene terminal connected with the superconductor terminal in the two-terminal junction, the reflected hole can only flow back into the same terminal as that of the incident electron. It is quite clear that the antichiral edge states due to the broken time-reversal symmetry play a key role in Andreev reflections. However, it is still difficult to present a clear physical illustration about the distinct role of antichiral edge states.

To find the answer, we choose a four-terminal junction, and study how the two kinds of Andreev reflections are influenced by the antichiral edge states. As shown in Fig. \ref{fig1}(c), terminals 1 and 3 are chosen to be graphene zigzag ribbon, and terminals 2 and 4 are superconducting leads. Due to the antichiral edge states, the incident electron of the edge modes travels parallelly to the interface of the graphene ribbon and the superconductor terminals. Therefore, when an incident electron comes from terminal 1, the retroreflected  and  specular reflected hole should flow into terminals 1 and 3, respectively.

Setting the split hopping $t_{1}=0.6\Delta_{s}$, we calculate the Andreev reflection coefficients in the four-terminal junction shown in Fig. \ref{fig6}, where the Andreev reflection coefficients change with the incident energy $E$.  It can be seen that for the different $\varepsilon_{0}$,  both $T_{A,11}$ and $T_{A,13}$ are quite large when $|E|<\Delta_{s}$ and show peaks at the gap edge $|E|=\Delta_{s}$, which is in good accord with the theory \cite{Beenakker}. On the whole, the value of $T_{A,13}$ is larger than that of $T_{A,11}$ for $\varepsilon_{0}=0$, but it becomes smaller than $T_{A,11}$ with increasing $\varepsilon_{0}$. Taking the case of $\varepsilon_{0}=1.0\delta$ as an example, we can observe that the maximum of $T_{A,11}$ reaches up to $1.35$ whereas $T_{A,13}$ only takes $0.35$ for the same energy. In this sense, Andreev retroreflection dominates when the on-site energy $\varepsilon_{0}$ is shifted up or down.

In Fig. \ref{fig6}(a), the curves of $T_{A,11}$ and $T_{A,13}$ are symmetric about $E=0$. In the other three figures, $T_{A,13}$ always keeps symmetric about $E=0$, but $T_{A,11}$ is not symmetric about $E=0$. As far as we know, the coefficients of Andreev reflection are generally symmetrical to the incident energy $E$, and the asymmetry of $T_{A,11}$ or $T_{A,13}$ was never reported before. Obviously, this asymmetry of $T_{A,11}$ should be ascribed to the broken time-reversal symmetry or antichiral edge states. However, there are a few of issues worth clarification and discussion: One is how the antichiral edge states, which break time-reversal symmetry, result in the asymmetry of $T_{A,11}$, the other is why $T_{A,13}$ always keeps symmetric and $T_{A,11}$ does not show asymmetrical characteristic for the two-terminal junction as shown in Figs. \ref{fig3} and \ref{fig5}.

\begin{figure}[tbp]
\includegraphics[width=1.0\columnwidth,angle=0]{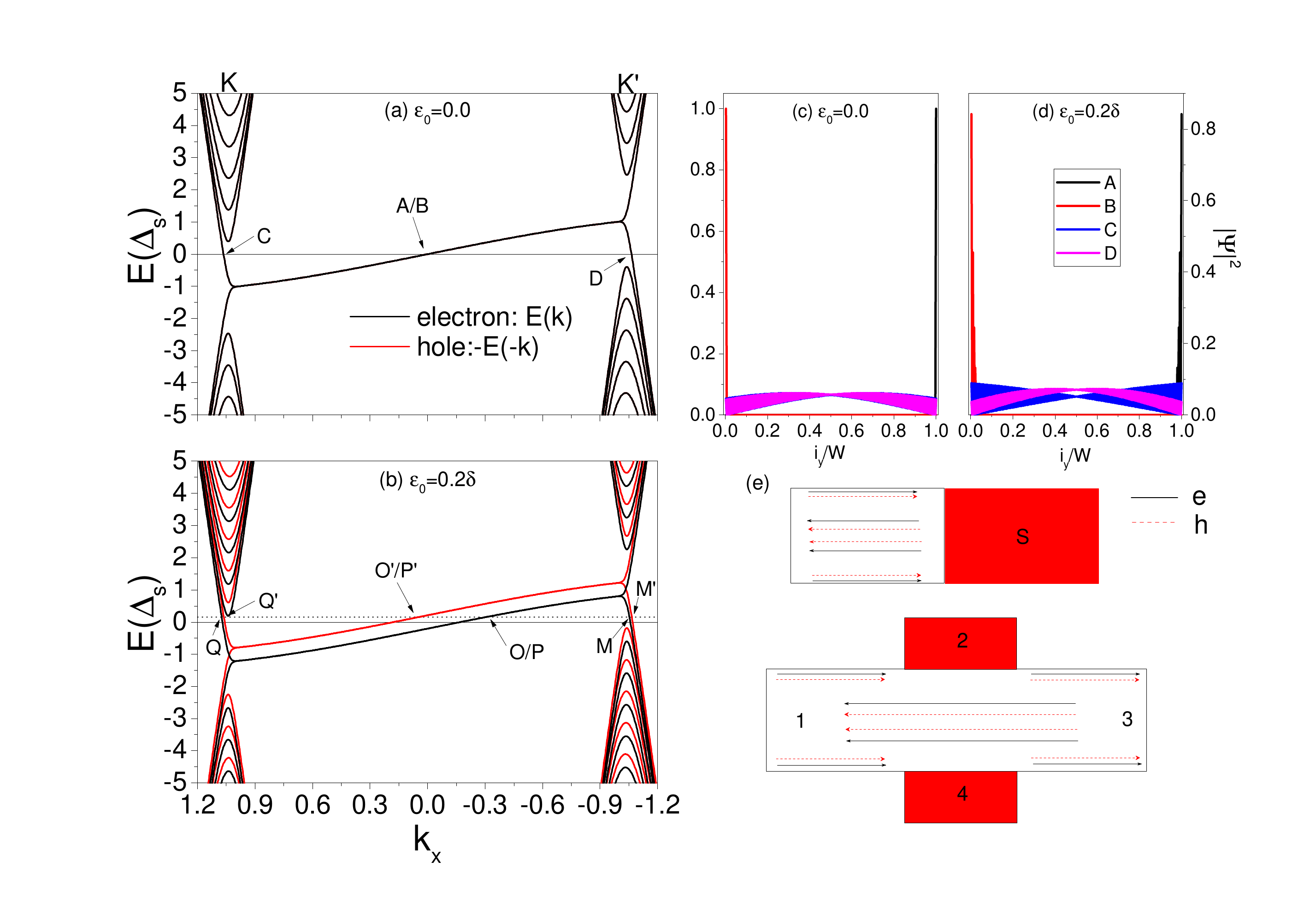}
\caption{(Color online) The band structures for both electron (black) and hole (red) at (a) $\varepsilon_0=0$ and (b) $\varepsilon_0=0.2\delta$. (c) and (d) represent the distribution of the electron states along the sample cross section. (e) The schematic diagrams of antichiral states in two-terminal and four-terminal junctions.} \label{fig7}
\end{figure}

To uncover the underlying physical reason, we plot the band structures for both electron (black) and hole (red) in Figs. \ref{fig7}(a) and 7(b), which relate to each other by electron-hole symmetry. As is seen, there are four electron states at Fermi energy in Fig. \ref{fig7}(a), which are denoted by states $A$, $B$, $C$, $D$, respectively. It is very much worth noting that all the hole states possess the same velocity as the corresponding electron states, due to the relationship of time-reversal symmetry between $K$ and $K'$ Dirac points. Undoubtedly, this peculiar characteristic would endow the antichiral states some unusual behaviours in Andreev reflections.  In addition, we scrutinize the distributions of the four electron states along the cross section of the sample in Figs. \ref{fig7}(c) and 7(d). Obviously, the states $A$ and $B$ with positive velocity are mainly located in two edges, which are therefore called antichiral edge states, while states $C$ and $D$ with negative velocity indeed correspond to bulk states. For the purposes of discussion below, in Fig. \ref{fig7}(e) we also present the schematic diagram of the movement of the states for two-terminal and four-terminal junctions, respectively.

Now, let us discuss why $T_{A,11}$ shows an asymmetrical feature in the four-terminal junction. From the calculation formula of Eqs. (\ref{AndRfC}), $T_{A,11}$ represents the probability of the Andreev reflection in which electrons from terminal $1$ are converted into holes of terminal $1$, and $T_{A,13}$ corresponds to the probability of the cross Andreev reflection in which electrons from terminal $1$ are converted into holes of terminal $3$. For $\varepsilon_0=0$, the properties of the initial electron and final hole states above the Fermi energy are the same as those of the corresponding initial electron and final hole states below the Fermi energy. Therefore, we observe symmetrical features of both $T_{A,11}$ and $T_{A,13}$ in Fig. \ref{fig6}(a).

The symmetry is destroyed when a nonzero value of $\varepsilon_0$ is taken. Note that the electron band (black) or the hole band (red) is no longer symmetrical to the incident energy $E=0$ because of a nonzero value of $\varepsilon_0$, as shown in Fig. \ref{fig7}(b). Due to electron-hole symmetry, an electron band (black) at the energy $E$ keeps the same to that of the corresponding hole state at the energy $-E$. For the sake of concreteness, we restrict to the case of $\varepsilon_0=0.2\delta$. In Fig. \ref{fig7}(b), there are also four electron states for a small energy ($E=|\eta|$) above the Fermi energy, represented by $O$, $P$, $Q$, $M$, respectively. If an incident electron from state $O$ is Andreev reflected as a hole of the state $O'$ or $P'$, it will contribute to the coefficient of $T_{A,13}$. While this incident electron is Andreev reflected as a hole of the state $Q'$ or $M'$, it will contribute to the coefficient of $T_{A,11}$. It is because both the initial electron state and the final hole state must move along the same direction for the cross Andreev reflection $T_{A,13}$, and the reflected hole must move oppositely as to the incident electron for the Andreev retroreflection $T_{A,11}$.

Although in the cross Andreev reflection, the initial electron state and the final hole state at the small positive energy $E=|\eta|$ are different from those at $E=-|\eta|$, they always keep conjugated with each other, namely the initial electron state and the final hole state at $E=|\eta|$ are equivalent to the final hole state and the initial electron state $E=-|\eta|$, respectively. From Fig. \ref{fig7}(b), we can see that the antichiral edge mode of the  electron is symmetric with that of the hole about the Fermi energy. It is not difficult to obtain that $T_{A,13}$ will maintain the symmetry about the Fermi energy even for a nonzero $\varepsilon_0$. But the symmetrical feature is destroyed in the Andreev retroreflection $T_{A,11}$. To sum up, the asymmetrical characteristic of $T_{A,11}$ in Figs. \ref{fig6}(b)-(d) should be ascribed to the coexistence of antichiral states and asymmetrical band structure.

All curves in Fig. \ref{fig6} become understandable by using the above-mentioned theory. In Fig. \ref{fig6}(a), $T_{A,13}$ shows a peak at small energy and decreases gradually with increasing the incident energy. However, $T_{A,11}$ keeps growing as the incident energy increases away from zero. It is because, the momentum separation between edge states and bulk states becomes  smaller with increasing the incident energy, which is conducive to $T_{A,11}$ but obstructive to $T_{A,13}$. For example, in Fig. 6(a), when $E=0$, the momentum difference between point A and D is biggest.  Therefore, when $E=0$, an electron of the edge states tends to be reflected as a hole of the edge states, which leads to the peak of $T_{A,13}$  and the valley of $T_{A,11}$ in Fig. 5(a). With $|E|$ increasing, the momentum difference becomes smaller and smaller, which leads to the decrease of $T_{A,13}$ and the increase of $T_{A,11}$. For a nonzero value of $\varepsilon_0$, the initial electron state is separated from the final hole state in momentum, as shown in Fig. \ref{fig7}(b), so $T_{A,13}$ becomes smaller and $T_{A,11}$ takes a nonzero value at $E=0$. As $\varepsilon_0$ takes a larger value, which leads to a large momentum separation between antichiral edge states of electron and hole  but a small one between electron edge states and hole bulk states, $T_{A,11}$ becomes dominated and $T_{A,13}$ is suppressed largely, as shown in Fig. \ref{fig6}(d). In a word, the variations of $T_{A,11}$ and $T_{A,13}$ in Fig. \ref{fig6} are mainly influenced by the magnitude of the transferred momentum in Andreev reflections.

There is still a serious question: Why $T_{A}$ (namely, $T_{A,11}$) keeps symmetrical in a two-terminal junction, distinctly different from the case of four-terminal junction. We can get hints from the schematic diagram of antichiral states in Fig. \ref{fig7}(e) and the band structure in Fig. \ref{fig7}(b). Obviously, for a small value of $\varepsilon_0$, there are two edge states with positive velocity and two bulk states with negative velocity, and the movement directions of corresponding hole states are the same as those of electron states. Therefore, the mutual scattering between edge and bulk states is inevitable when incident electrons from terminal $1$ are Andreev reflected as the holes of terminal $1$. Although the system has antichiral states and asymmetrical band structure with broken time-reversal symmetry, the abundant and sufficient scattering between antichiral edge and bulk states have balanced their influences on the Andreev retroreflection coefficient $T_{A,11}$. So $T_{A,11}$ manifests a symmetrical feature about the incident energy.

\begin{figure}[tbp]
\includegraphics[width=1.0\columnwidth,angle=0]{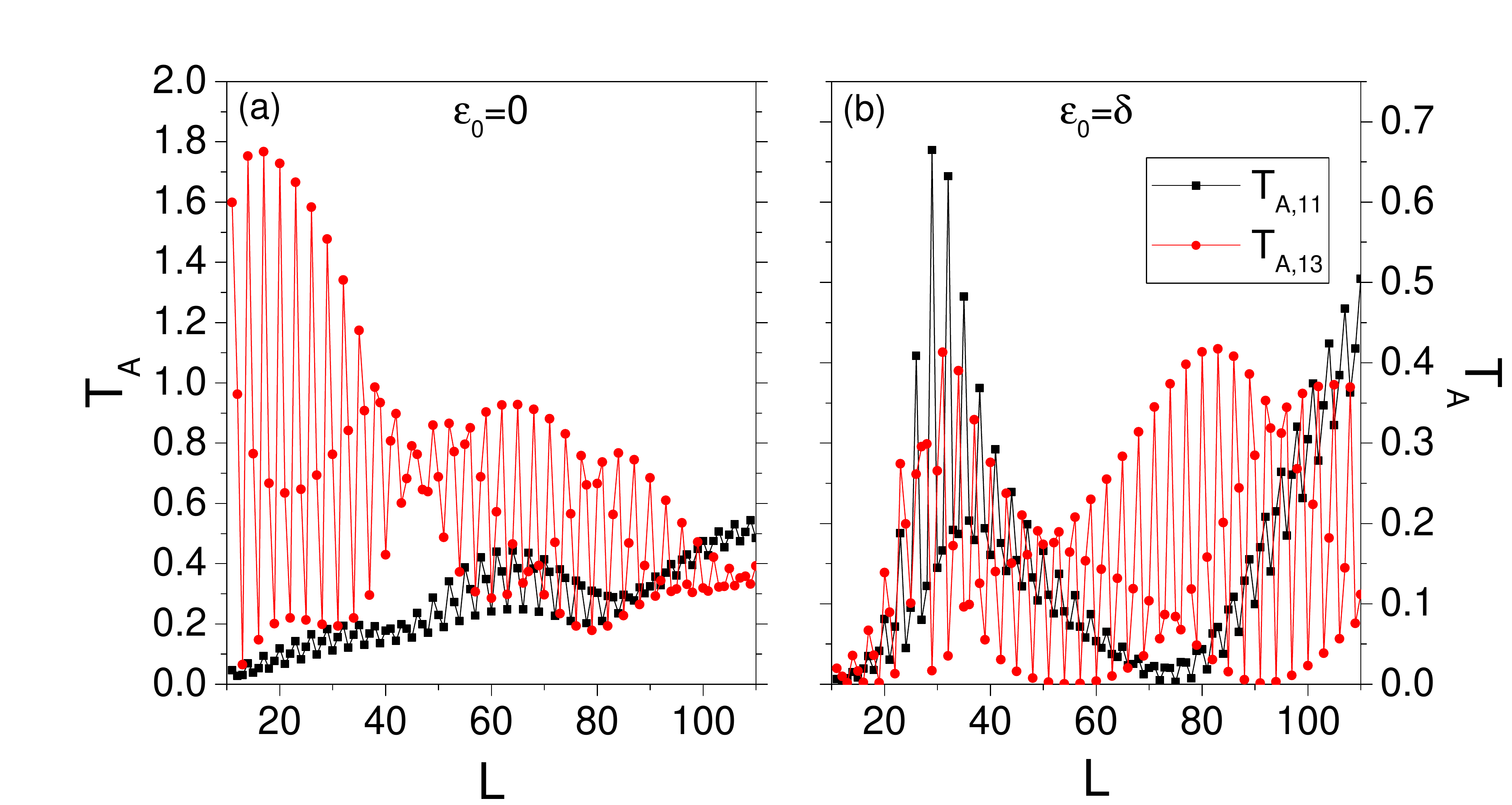}
\caption{(Color online) $T_{A,11}$ and  $T_{A,13}$ vs the length L of the center region  in the four-terminal graphene-superconductor junction.  $\varepsilon_{0}$ takes $0$ and $\delta$ in (a) and (b), respectively. Here, the split strength $t_{1}=0.6\Delta_{s}$, the width $W=80$, and the incident energy $E=0.01\Delta_{s}$. } \label{fig8}
\end{figure}

At last, Fig. \ref{fig8} shows the coefficients of Andreev reflection in the four-terminal junction as the function of the length $L$. The incident energy is set to be small enough to be close to the Fermi energy. With $L$ increasing, there are multiple Andreev reflections occurring between two interfaces. Therefore, both $T_{A,11}$ and $T_{A,13}$ oscillate intensely with $L$ increasing no matter what the value $\varepsilon_{0}$ takes. Comparing the curves of $T_{A,11}$ and $T_{A,13}$ in Fig. \ref{fig8}(a), it can be seen that the amplitude of $T_{A,13}$ is bigger than that of $T_{A,11}$. Especially, there is one distinct difference between two curves in Fig. \ref{fig8}(a). When $L$ is small enough, $T_{A,13}$ can reach up to $1.8$, which is substantially larger than $T_{A,11}$ in Fig. \ref{fig8}(a). However, both $T_{A,11}$ and $T_{A,13}$ are close to zero for small $L$ in Fig. \ref{fig8}(b). It is easy to understand by analyzing the band structures at $\varepsilon_0=0$ and $\varepsilon_0=\delta$. In the case of $\varepsilon_0=0$, there are antichiral edge states at $E=0.01\Delta_s$, so the incident electron from terminal $1$ is apt to be specularly reflected to terminal 3. Whereas normal bulk states are predominant at $E=0.01\Delta_s$ in the case of $\varepsilon_0=\delta$, incident electron is probable to tunnel directly into terminal 3, which leads to very small values for both $T_{A,11}$ and $T_{A,13}$. As increasing the length $L$ over $80$, in Fig. \ref{fig8}(b), $T_{A,11}$ and $T_{A,13}$ increase gradually due to the growing contribution of bulk states to Andreev reflection.

\section{Conclusions}
We study the Andreev reflection in a zigzag graphene ribbon with split Dirac cones. Due to the antichiral edge modes with the same velocity and dispersion, the time-reversal symmetry in a pristine graphene model is broken up. Different from the pristine graphene model, the incident electrons of the antichiral edge states can make an obvious contribution on Andreev reflection.

In a two-terminal graphene-superconductor junction, the Andreev reflection coefficient $T_{A}$ takes nonzero value within the range of $|E|<(\delta-\varepsilon_{0})$, where $\delta=\sqrt{3}t_1$ represents the energy difference between the present Dirac cones and the pristine ones. Especially, $T_{A}$ maintains a symmetrical feature about the incident energy, as reported in previous papers. It is worth noting that the strength of $T_{A}$ can be tuned by changing the on-site energy $\varepsilon_0$.

Different from the case of the two-terminal junction, in a four-terminal junction the coefficient of Andreev reflection $T_{A,13}$ shows a symmetrical feature about the incident energy but $T_{A,11}$ manifests an asymmetrical characteristic. Through analysis, this distinct characteristic should be ascribed to the coexistence of antichiral states and asymmetrical band structure. Note that there should be an abundant and sufficient scattering between antichiral edge and bulk states for Andreev retroreflections of a two-terminal junction, which helps $T_{A,11}$ to get rid of the influence of broken time-reversal symmetry and preserve the symmetrical characteristic. This is why the Andreev reflection coefficient $T_{A,11}$ in the two-terminal junction always keeps symmetrical to the incident energy. Besides the effect of the structure,  the reasons for symmetric $T_{A,11}$ in the two-terminal junction are complicated, so many researching works are needed to clarify the underlying physics.

The results in this paper are very important to understand intervalley reflection, intravalley reflection, interband reflection, and intraband reflection and helpful to exploit the graphene-superconductor junction. In addition, this paper presents a clear physical picture about the behaviours of antichiral states in Andreev reflection. It could be important to find new materials and functional quantum devices.  In this paper, we focus on the Andreev reflection under the BCS mechanism. However, we have noted that the FFLO mechanism ~\cite{Ferrell,Ovchinnikov} should be a significant question, which could also have influence on the Andreev reflection. It is an open question worth studying in further research.

\section*{ACKNOWLEDGMENTS}

This work was supported by the National Natural Science Foundation of China (Grants No. 11874139 and No. 11474085), the Natural Science Foundation of Hebei (Grants No. A2017205108 and No. A2019205190), the youth talent support program of the Hebei education department (Grant No. BJ2014038), the Outstanding Youth Foundation of HBTU (Grant No. L2016J01), and the youth talent support program of Hebei Province.\\

\end{document}